\documentclass[reprint, amsmath,amssymb, aps, prl, longbibliography, floatfix]{revtex4-1}

\usepackage{graphicx}% Include figure files
\usepackage{dcolumn}% Align table columns on decimal point

\usepackage{bm}% bold math
\usepackage{hyperref}% add hypertext capabilities
\usepackage{braket}
\usepackage{longtable}
\usepackage[usenames, dvipsnames]{color}
\usepackage[mathlines]{lineno}% Enable numbering of text and display math
\usepackage{fancyhdr}
\definecolor{mygray}{gray}{0.6}

\begin{document}

%\preprint{APS/123-QED}

\title{Superconducting Caps for Quantum Integrated Circuits}

\author{William O'Brien, Mehrnoosh Vahidpour, Jon Tyler Whyland, Joel Angeles, Jayss Marshall, Diego Scarabelli, Genya Crossman, Kamal Yadav, Yuvraj Mohan, Catvu Bui, Vijay Rawat, Russ Renzas, Nagesh Vodrahalli, Andrew Bestwick, Chad Rigetti}

\affiliation{Rigetti Computing, 775  Heinz  Avenue,  Berkeley,  CA  94710}

%\collaboration{Rigetti Computing}%\noaffiliation

\date{\today}
%\date{June 20, 2017}

\begin{abstract}
We report on the fabrication and metrology of superconducting caps for qubit circuits. As part of a 3D quantum integrated circuit architecture, a cap chip forms the upper half of an enclosure that provides isolation, increases vacuum participation ratio, and improves performance of individual resonant elements. Here, we demonstrate that such caps can be reliably fabricated, placed on a circuit chip, and form superconducting connections to the circuit.
\end{abstract}

\maketitle

%%%%%%%%%%%%%%%%%%%%%%%%%%%%%%%%%%%%%%%%%%%%%%%%%%%%%%%%%%%%%%%%%%%%%%%%%%%%%%%%%%%%%%%%%%%%%%%%%%%%%%%%%%%%%%%%
\section{Motivation}

Superconducting qubits are highly sensitive detectors of material quality. The coupling of qubits' electromagnetic modes with lossy, defective materials can limit their coherence time. In a 2D circuit geometry the field lines are largely in the plane, as in Fig. 1A, where the mode interacts strongly with interface and substrate defects. One way to avoid this is to place qubits inside a 3D cavity, which increases the modes' participation in lossless vacuum, thereby reducing microwave loss and increasing qubit lifetime~\cite{3Dtransmon}. However, this approach sacrifices the scalability and density of 2D circuits.

\begin{figure}
%\rule{0.45\textwidth}{4cm}
\includegraphics[width=0.45\textwidth]{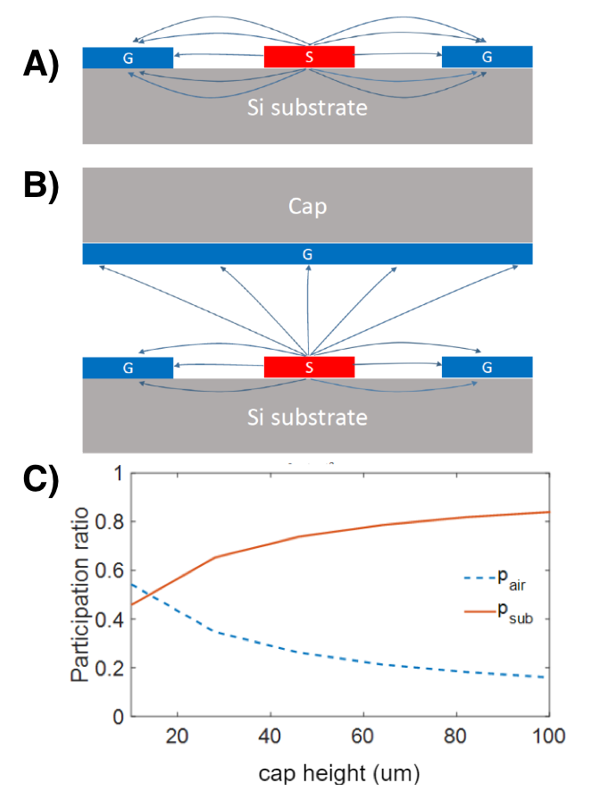}
\caption{Schematic illustration of cap effect on mode profile. Arrows indicate how the electric field lines, extending from source (S) to ground (G), change in a A) 2D and B) 3D circuit geometry. It can be seen that the 3D geometry pulls more of the field lines into free space, decreasing dielectric losses. C) Modeling results show the expected vacuum participation enhancement with decreasing cap height. P$_{air}$ and P$_{substrate}$ are the fraction of electric fields in the vacuum and substrate, respectively.}
\label{CavEffect}
\end{figure}

 To confer some benefits 3D cavities onto a 2D circuit, we propose a geometry as illustrated in Fig.~\ref{CavEffect}B in which a superconducting cap is placed on top of the chip. Here, field lines terminate preferentially on the walls of the enclosure, which increases the spatial overlap of the mode with the lossless medium of free space. Microwave simulations of superconducting resonators underneath a cap, as in Fig.~\ref{CavEffect}C., indicate a strong response of the electromagnetic vacuum participation ratio to the cap height.

In addition, when employing a non-planar quantum integrated circuit architecture including superconducting vias, we can engineer separate shielded enclosures to improve the electromagnetic isolation and reduce crosstalk between circuit elements. Simulations indicate that such enclosures reduce crosstalk between neighboring resonant elements by 15 dB.

In summary, we expect that bonding caps with a superconducting liner over qubit circuits will confer improved coherence times, reduced crosstalk, and better immunity to environmental noise.

\section{Fabrication}
\begin{figure}
%\rule{0.45\textwidth}{4cm}
\includegraphics[width=0.5\textwidth]{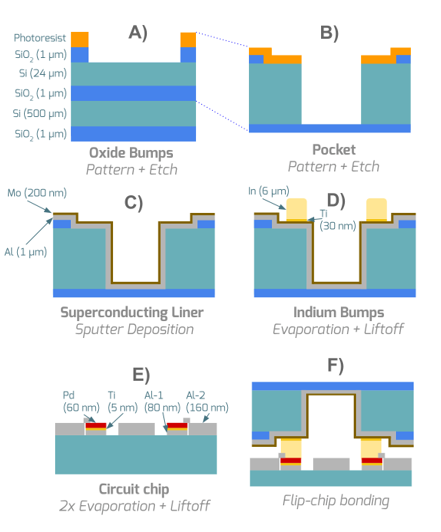}
\caption{Process flow diagram for cap fabrication. A) Oxide bumps patterned and etched, B) DRIE pocket etched using Bosch process through photoresist mask down to oxide stop, C) sputter deposition of Al/Mo, D) indium bumps with titanium adhesion layer deposited by e-beam evaporation, after liftoff process, E) circuit chip fabrication, involving Al/Ti/Pd pads for bonding to indium, F) final bonded structure.
\label{fabprocess}}
\end{figure}

\begin{figure}
%\rule{0.45\textwidth}{4cm}
\includegraphics[width=0.5\textwidth]{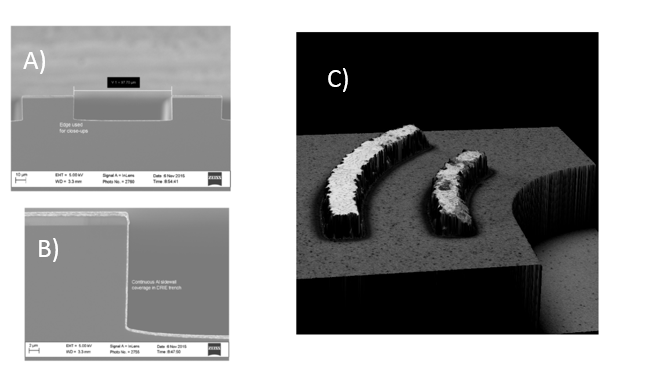}
\caption{Images from cap fabrication. A) and B) are SEM images after sputter deposition, displaying smooth, vertical sidewalls and a continuous superconducting liner. Shown in C) is a confocal microscope image of the profile after liftoff of the indium bumps.
\label{CapSEM}}
\end{figure}

The cap is composed of pockets etched in Si, coated with superconducting metal, and patterned with indium bumps as a superconducting adhesive for bonding to the circuit chip. Caps are fabricated from silicon-on-insulator wafers, in which a 24 $\mu$m Si device layer is sandwiched on both sides by 1 $\mu$m of oxide, with thick Si underneath for structural support. The top side of the wafer is patterned with photoresist, first to mask an inductively coupled plasma etch that selectively removes the top oxide (Fig.~\ref{fabprocess}A) to pattern bumps that limit the gap between cap and circuit, and then a second time to mask a deep reactive ion etch (DRIE) down to the buried oxide. The DRIE follows the Bosch process~\cite{bosch}, producing a square pocket with vertical sidewalls 24 $\mu$m deep (Fig.~\ref{fabprocess}B). The DRIE step employs a LF substrate bias to avoid footing near the buried oxide. The surface is then conformally coated with 1 $\mu$m sputtered Al to form a continuous superconducting shield. The resulting profile is visualized in Fig.~\ref{CapSEM}A and Fig.~\ref{CapSEM}B, confirming a continuous sputtered film and smooth, vertical sidewalls from the Bosch process.

In order to form high quality interfaces, the terminating metal layers on the cap and circuit (which will mediate the indium connection) are selected to avoid or mitigate native oxides. On the cap side, termination with an Al native oxide is avoided by capping the Al in-situ (without breaking vacuum) with a thin, 200 nm Mo film (Fig.~\ref{fabprocess}C). Mo is a refractory metal that is superconducting near 1 K. Similarly, Pd is chosen to terminate bond pads on the circuit side. Although Pd is not superconducting at our operating temperature, it is designed to be thin enough to still behave as a superconductor through the proximity effect.

To form this electrical and mechanical connection between circuit and cap, evaporated indium bumps (with Ti adhesion layers) are patterned on the cap. Electrically, indium serves as an ideal Type-I superconducting bond to ground the cap, as it has a $T_c$ above 3 K. Several groups have successfully used superconducting indium bumps to bond chips with qubits to another chip with readout and control signals with high qubit coherence ($>$20 $\mu$s) \cite{lincoln, google,google2}, which is a more technically demanding application. Electron-beam evaporation is chosen for indium deposition, since it is well known to produce high quality indium films, and is amenable for liftoff (Fig.~\ref{fabprocess}D). The patterned bumps are found to be well defined, as seen in the confocal microscope image of Fig.~\ref{CapSEM}C.

The circuit chip is produced in two liftoff steps. First, we pattern bond pads for contacting the indium, deposit 85 nm Al, 5 nm Ti, and 60 nm Pd, then perform liftoff. The palladium is a noble metal that does not oxidize, and is commonly used for bonding purposes. Although it does not superconduct, it will experience the superconducting proximity effect by being sandwiched between Al and In. It is in this way possible to form a continuous superconducting path from the circuit to cap chip despite the use of normal metals, albeit it with a weak link, so long as those normal layers are thin enough. The second step of the circuit chip fabrication is to pattern, deposit, and liftoff superconducting circuit features (ground plane, resonators, etc.) with 160 nm of Al, allowing for some overlap on the bonding pads to ensure a good electrical connection (Fig.~\ref{fabprocess}E). This is performed in a Plassys e-beam evaporation system, designed for Josephson Junction (JJ) fabrication.

To form the 3D structure, we use a flip-chip bonder to affix the cap chip to the circuit chip (Fig.~\ref{fabprocess}F). The two chips are precisely aligned and pressed together under up to 45 kg of weight at around 70$^\circ$ C. Note that the 1 $\mu$m oxide bumps are patterned to place a limit on how close the cap and circuit chip come in contact during bonding. Though in practice, we do not need to apply the required force to press the indium flush with the oxide bumps.

\section{Test structures}

\begin{figure}
%\rule{0.45\textwidth}{4cm}
\includegraphics[width=0.45\textwidth]{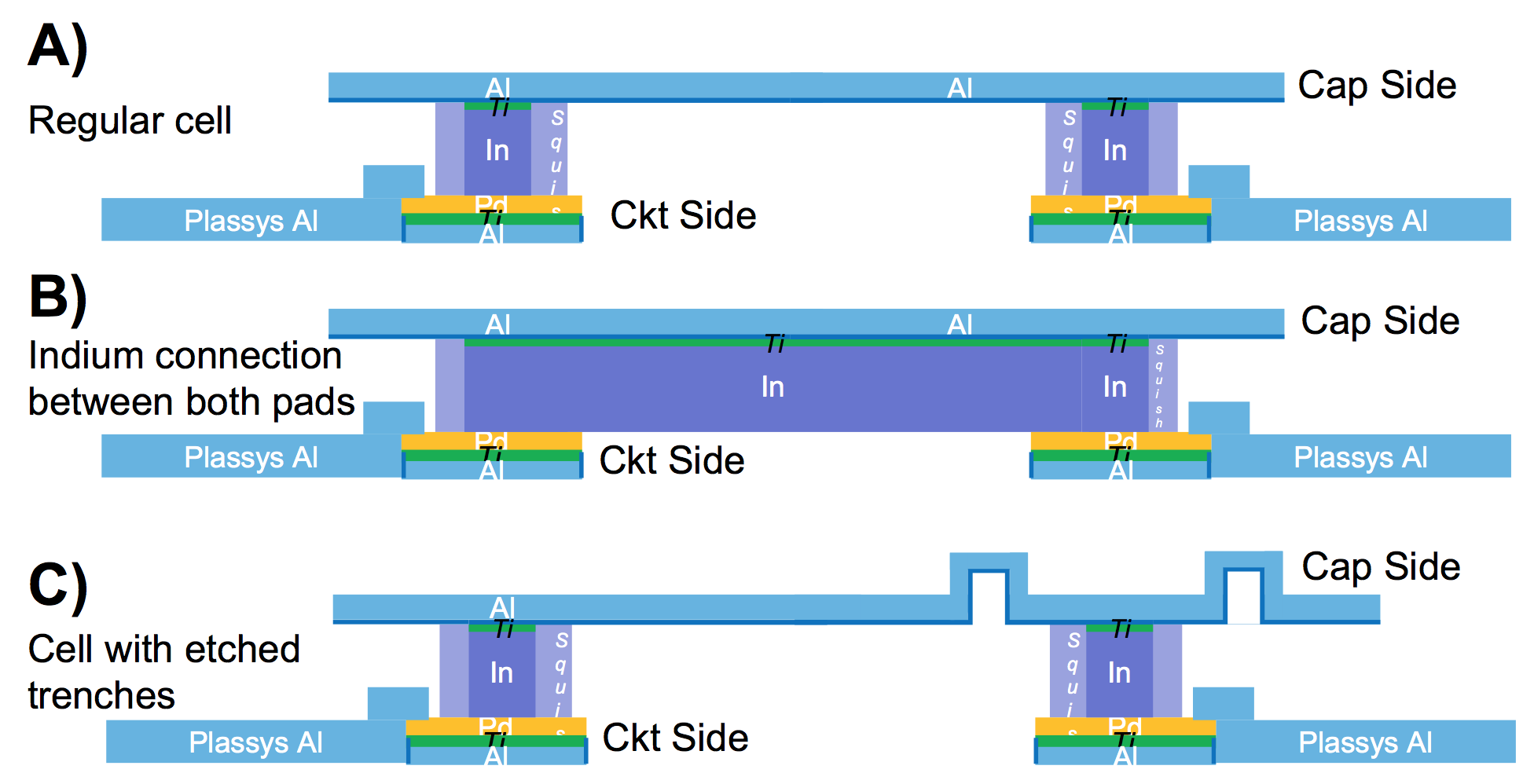}
\caption{Schematic of test structures for indium bonding, in profile. A) Regular cells compose nearly 90$\%$ of the sites, and include structures to measure the roundtrip resistance from circuit to cap and back, including the electrical contact in the full bond stack Pd-In-Ti-Mo-Al. B) Indium shorts out adjacent pads, which isolates the impact of the In-Pd interface. C) Trenches are etched around adjacent indium bumps, which could potentially sever continuity if the Al film is not entirely conformal.
\label{stackup}}
\end{figure}

To assess the quality of the bond across the die, test structures shown in Fig.~\ref{stackup} were designed to measure series resistance across indium bumps. The conductive path is designed such that current must traverse from the circuit metal to the cap through an indium bump, then back to the circuit along an adjacent indium bump. Some special test structures were also implemented, to test the conformality of the sputtered Al, and others to test the intrinsic contact resistance between indium and palladium. To test for Al conformality, trenches with the same depth as the pocket are etched into the cap between adjacent indium bumps. If the Al layer is not continuous, current injected through one bump is not able to find a return path to the circuit. To test the contact resistance of just the indium-palladium interface, two adjacent pads are shorted with indium, such that current does not need to flow through the cap for continuity.

\begin{figure}
\includegraphics[width=0.45\textwidth]{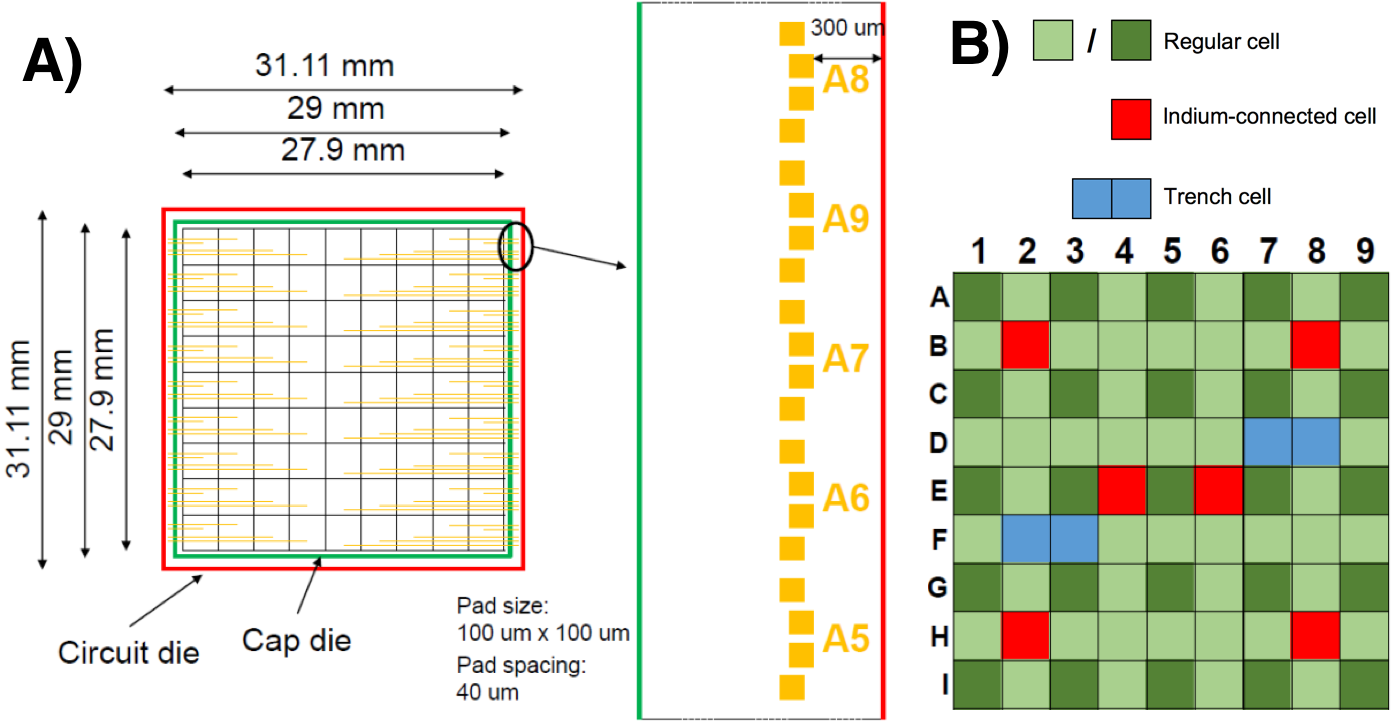}
\caption{A) Arrangement of 81 test structures on a 31 mm circuit chip and 29 mm cap chip. Bond pads are at the perimeter of the circuit chip to allow for probing after bonding. B) Distribution of test structure types across chip.
\label{TestStructures}}
\end{figure}

The test structures were fabricated on a 31 mm circuit chip and a corresponding 29 mm cap chip, arranged in a $9 \times 9$ grid with signal lines designed so that all 81 structures could be measured from the circuit chip edges (Fig.~\ref{TestStructures}A). This arrangement allows us to probe sites distributed across a large chip and test planarity of the bond. Most of the test structures were standard cells, with 6 In-connected cells and 4 with trenches (Fig.~\ref{TestStructures}B).

\section{Results}

At room temperature, series resistances below 5 $\Omega$ were obtained at 80 out of 81 test sites, demonstrating excellent continuity and high process fidelity. From the lowest resistance values measured, the room temperature contact resistance between indium and molybdenum is less than 0.2 $\Omega$. No difference was measured between the different test structure types, indicating that the Al layer is indeed continuous across the pockets. Furthermore no major asymmetries in resistance were observed between the left and right sides or top and bottom of the chip, indicating that bond parallelism across a large 31 mm chip is acceptable.
\begin{figure}
\includegraphics[width=0.45\textwidth]{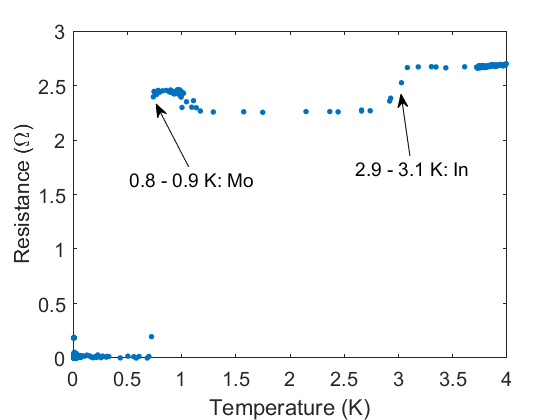}
\caption{Cryogenic DC resistance with labeled superconducting transitions associated with Mo/Al and In. The instrumentation used for this measurement sets a lower bound for the critical current of 36 $\mu$A.
\label{CryoResistance}}
\end{figure}

We also performed cryogenic measurements to assess the superconducting properties. An example is shown in Fig.~\ref{CryoResistance}. As we cool the sample down, we see sharp resistance drops at 2.9-3.1 K and 0.8-0.9 K, which correspond approximately to the superconducting transition temperatures of In and Mo/Al, respectively. The absence of separate Al and Mo transitions could be explained by a variety of superconducting-normal interface effects.

In any case, under 0.8 K we observe zero resistance, below the sensitivity of our instruments, indicative of a superconducting path from the circuit chip to the cap chip and back. Our instruments set the lower bound of the critical current of this path as 36 $\mu$A.

\section{Conclusions}
The application of a Mo capping layer to the Al cap has proven to be an effective means of mitigating the deleterious effects of native oxide on the electrical and mechanical contact of indium bumps between circuit and cap die. In addition, the  consistently low room-temperature resistances and clear low-temperature superconductivity demonstrate that the cap can be kept grounded by robust supercurrents through the bonds, at locations across the die.

With successful basic metrology of this technology established, we can move onto integration with superconducting quantum circuits, including qubits. Further work is needed to systematically establish enhancements to qubit performance.

\end{document}